\documentclass[a4paper]{article}

\usepackage[numbers]{natbib}
\usepackage{rotating}
\usepackage{graphicx}
\usepackage{subfig}
\usepackage{amssymb}
\usepackage{authblk}

\newcommand{\pt}{p_\mathrm{T}}

\newcommand{\runi}{Run I}
\newcommand{\runii}{Run II}

\newcommand{\pbpb}{Pb--Pb}

\begin{document}

\title{The ALICE Transition Radiation Detector:\\
  status and perspectives for \runii{}}

\author[1]{Jochen Klein\thanks{Jochen.Klein@cern.ch}\\
  for the ALICE Collaboration}
%
\affil[1]{CERN}

\date{}

\maketitle

\begin{abstract}
The ALICE Transition Radiation Detector contributes to the tracking,
particle identification, and triggering capabilities of the
experiment. It is composed of six layers of multi-wire proportional
chambers, each of which is preceded by a radiator and a
Xe/CO$_2$-filled drift volume. The signal is sampled in timebins of
100~ns over the drift length which allows for the reconstruction of
chamber-wise track segments, both online and offline. The particle
identification is based on the specific energy loss of charged
particles and additional transition radiation photons, the latter
being a signature for electrons.

The detector is segmented into 18 sectors, of which 13 were installed
in Run I. The TRD was included in data taking since the LHC start-up
and was successfully used for electron identification and triggering.
During the Long Shutdown 1, the detector was completed and now covers
the full azimuthal acceptance. Furthermore, the readout and trigger
components were upgraded. When data taking was started for \runii{},
their performance fulfilled the expectations.
\end{abstract}

\section{INTRODUCTION}

A Large Ion Collider Experiment (ALICE) is the experiment at the Large
Hadron Collider (LHC) at CERN which was particularly optimized for the
measurement of \pbpb{} collisions~\cite{pap:alice}. It consists of a
central barrel and a forward muon spectrometer. The former is located
in a warm magnet providing a solenoidal field of $B = 0.5~\mathrm{T}$.
A large cylindrical Time Projection Chamber (TPC) is used as the main
tracking device. It is complemented by a silicon-based Inner Tracking
System (ITS) close to the beam pipe and, towards larger radii, the
Transition Radiation Detector (TRD) and the Time-Of-Flight (TOF)
detectors. Part of the acceptance is covered by electromagnetic
calorimetry, and further detectors are installed around the
interaction point for triggering and event characterization.

Following the azimuthal segmentation of the ALICE central barrel, the
TRD is organized in 18~sectors~\cite{tdr:trd}. Each of them is filled
with a supermodule consisting of 6~layers of Multi-Wire Proportional
Chambers (MWPC). They are subdivided into five stacks to achieve
manageable chamber sizes. The MWPCs are preceded by a drift volume and
a fibre-foam radiator. The former allows for the detection of the
ionization energy loss over a radial length of $3.7~\mathrm{cm}$. The
detection includes the absorption of the transition radiation, which
highly relativistic ($\gamma \gtrsim 800$) particles can emit while
traversing the radiator. The transition radiation photons are
predominantly in the X-ray regime, and Xenon is used as detection gas
because of its high photon absorption cross section. For the central
barrel, the TRD contributes tracking, triggering, and the
identification of particles, in particular of electrons.

In the following, we will first discuss the setup and operation of the
TRD in \runi{}. We will further summarize results showing the
performance of the detector. Next, we will describe the consolidation
and upgrade activities during the first long shutdown of the LHC, as
well as the subsequent recommissioning. In the end, we will review the
current situation and further plans for \runii{}.

\section{RUN I}

\begin{figure}
  \centering
  \captionsetup{type=figure}
  \subfloat[ALICE central barrel (end of \runi{})]{
      \includegraphics[height=.37\textwidth]{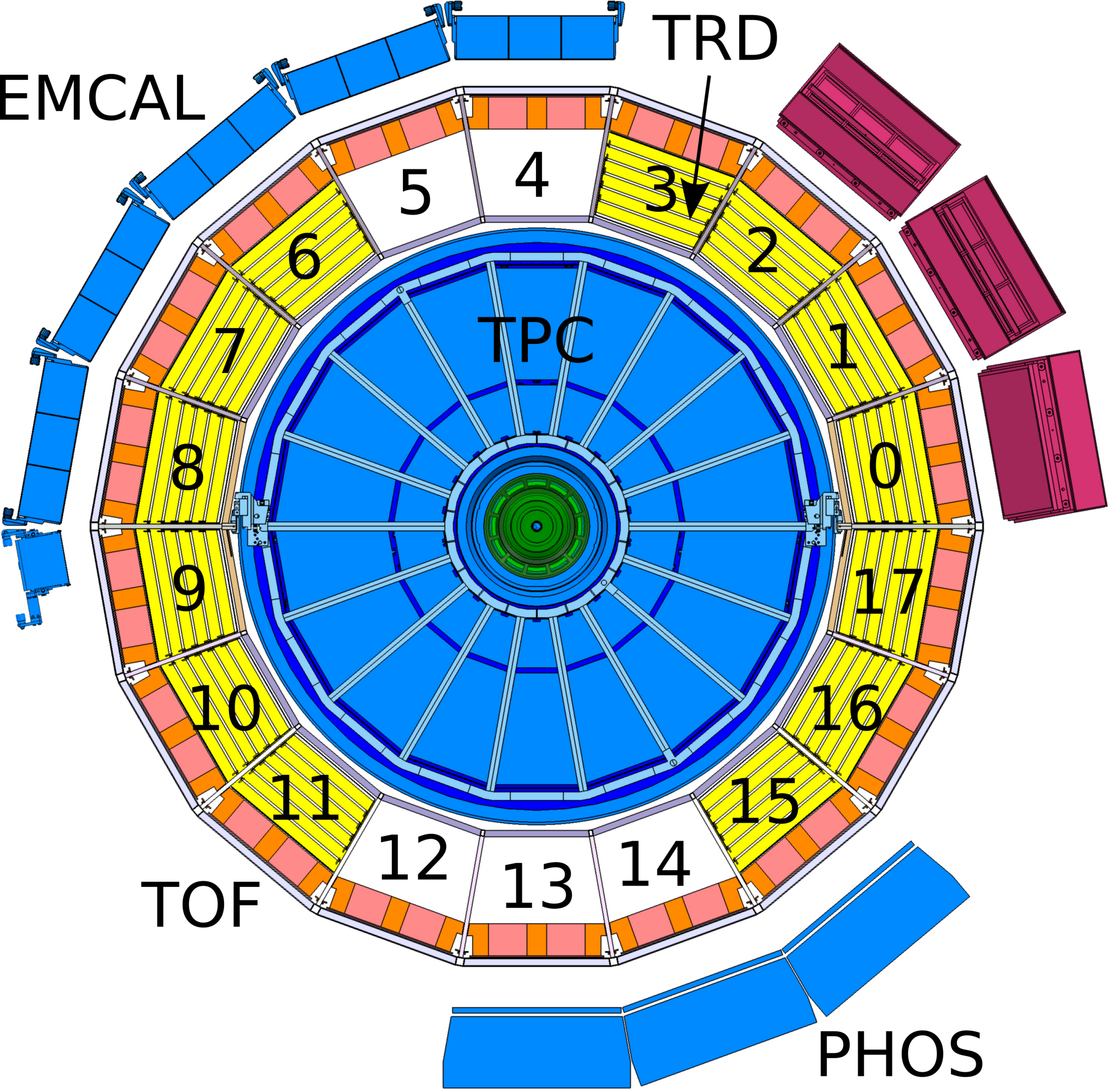}
  }
  \hspace{.5cm}
  \subfloat[energy deposit in the TRD]{
      \includegraphics[height=.37\textwidth]{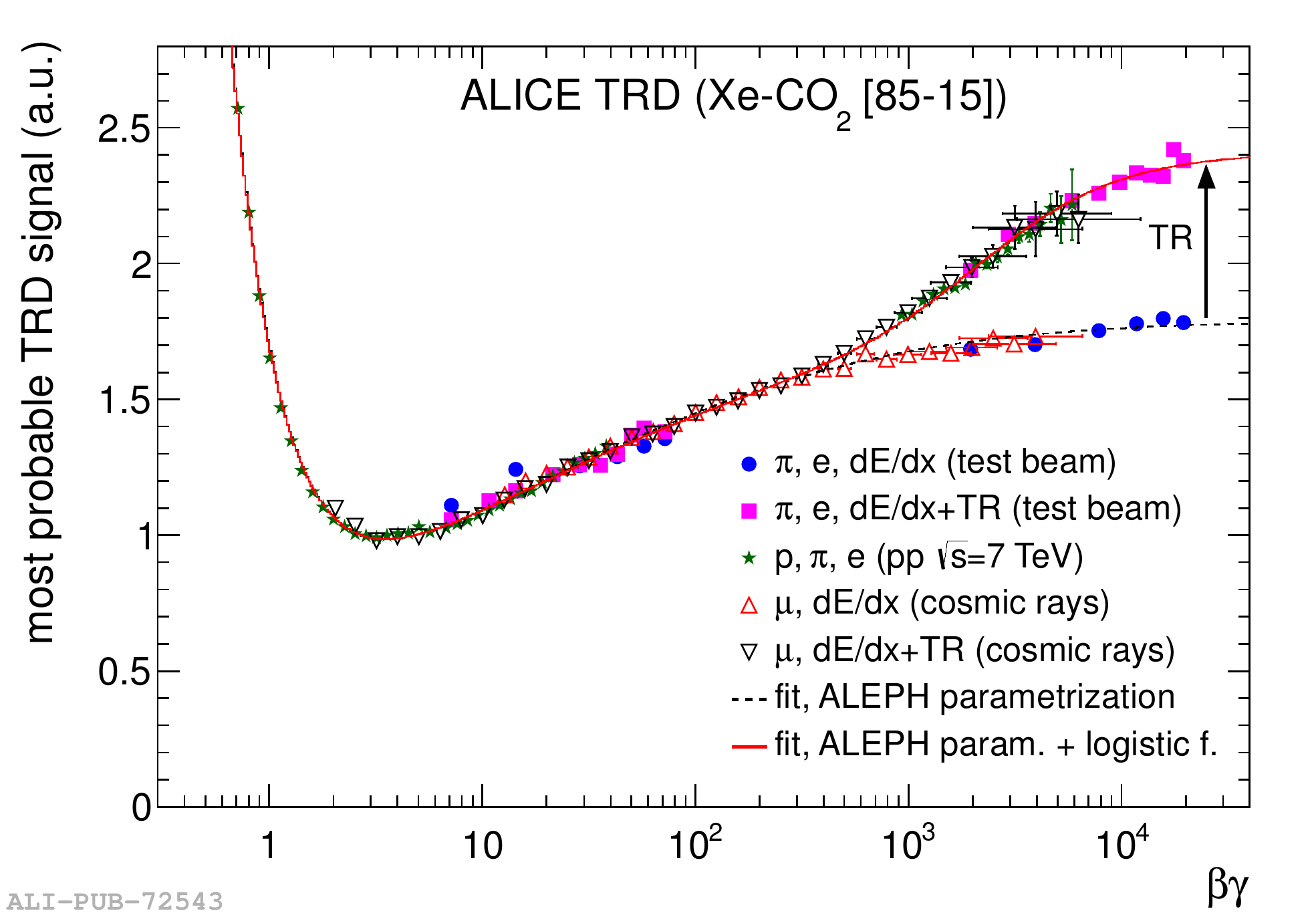}
  }
  \caption{TRD in the ALICE central barrel. Left: Cross-sectional view
    with the installation status of 2012/13. Right: Energy deposit in
    the TRD with and without transition radiation.}
  \label{fig:alice}
\end{figure}

Since the beginning of \runi{}, the TRD was included in the data
taking of ALICE with the supermodules already installed at that
time~\cite{pap:alice_perf}. By 2012, 13 out of 18 had been installed,
resulting in the setup shown in Figure~\ref{fig:alice}.

The central barrel tracking is based on a Kalman filtering
approach~\cite{pap:alice_perf}. Starting from seeding clusters at the
outer radius of the TPC, tracks are found by inward propagation to the
vertex, also attaching hits in the Inner Tracking System. In a second
step, the track is propagated outwards through the TRD and TOF
detectors, also attaching clusters there. In a final inward
propagation, the track parameters are refitted, taking into account
energy loss. The TRD clusters carry both position and charge
information, the latter allowing for the particle identification.

The fundamental concept, i.e. the detection of specific energy loss
and the onset of Transition Radiation (TR), could be verified using
cosmic muons~\cite{th:lu}. They can carry very large momenta and can
traverse the TRD twice. In the outward direction, which corresponds to
the normal scenario of particles coming from the interaction point,
the radiator is traversed before the chamber and the transition
radiation emitted into the drift volume. On the inward direction,
however, the radiator is passed only after the chamber and the
transitition radiation remains undetected. This allows for the
separation of the measurements with and without transition radiation.
Figure~\ref{fig:alice} shows a compilation of measurements in the
experiments and from previous test beams. The cosmic muons bridge the
gap between data points from test beam setups with electron and pion
samples and confirm the expected onset of transition radiation around
$\beta \gamma \simeq 800$.

\begin{figure}
  \centering
  \captionsetup{type=figure}
  \subfloat[layer dependence]{
    \begin{minipage}[c][1\width]{.45\textwidth}
      \centering
      \includegraphics[width=\textwidth]{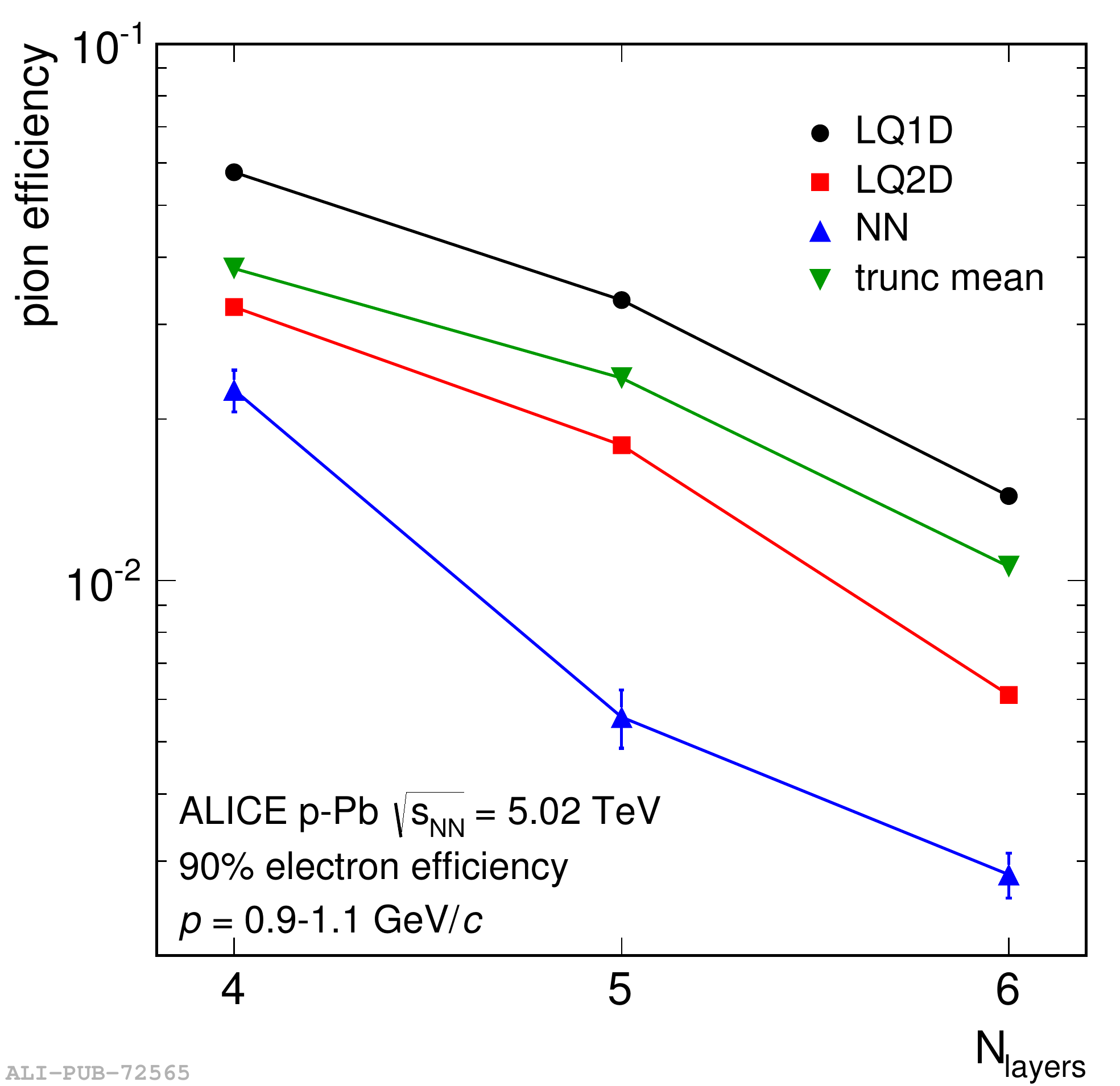}
    \end{minipage}
  }
  \hspace{.5cm}
  \subfloat[momentum dependence]{
    \begin{minipage}[c][1\width]{.45\textwidth}
      \centering
      \includegraphics[width=\textwidth]{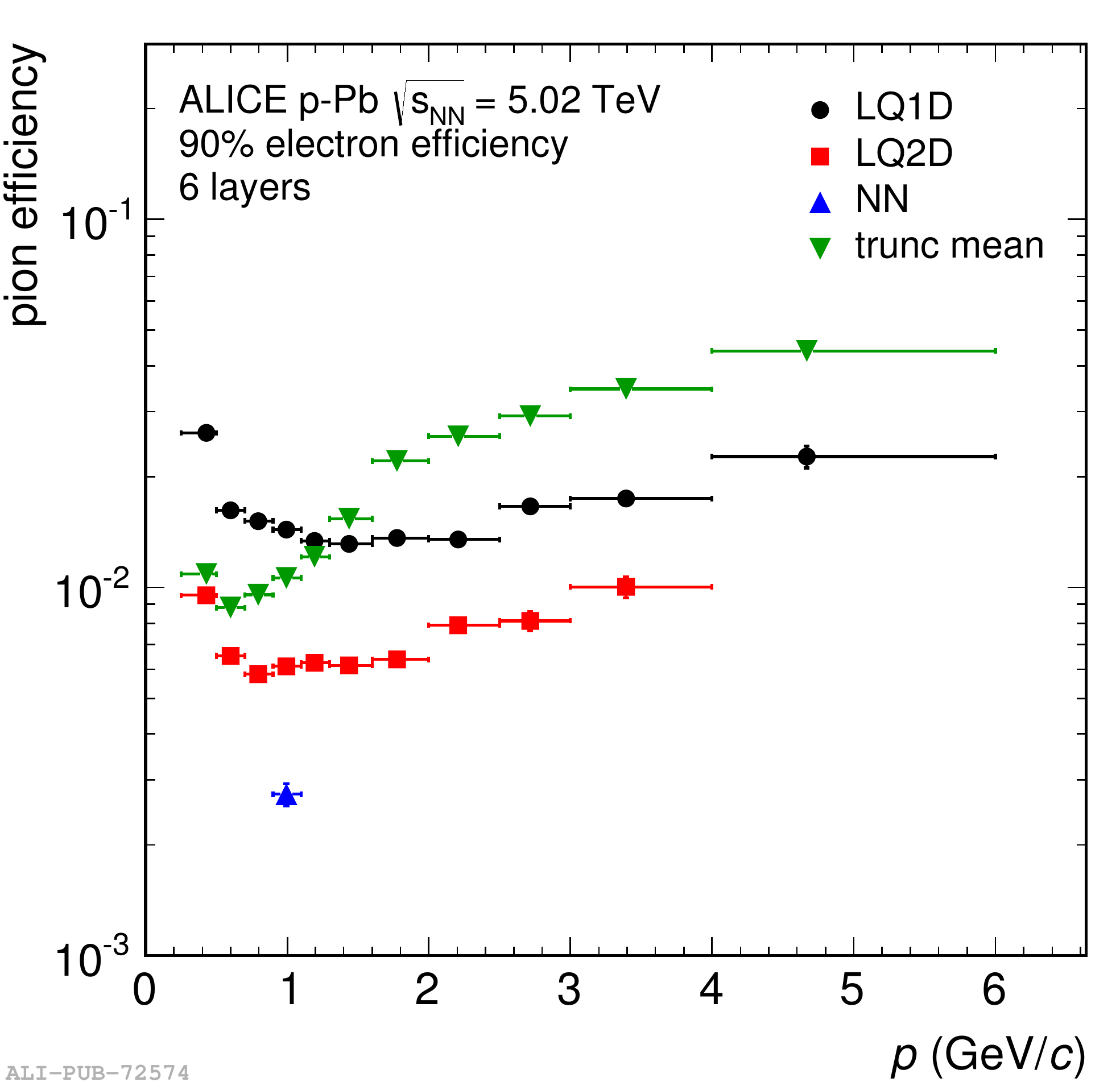}
    \end{minipage}
  }
  \caption{Electron identification with the TRD~\cite{pap:alice_perf}.
    The efficiency (inverse of rejection) for pions at a given
    electron efficiency of 90\% is shown.}
  \label{fig:eid}
\end{figure}

For electron identification, a likelihood can be calculated based on
the total accumulated charge, which comprises the energy loss from
ionization in the active volume and, if present, the absorption of
transition radiation~\cite{th:lohner}. The transition radiation is
most likely absorbed close to the entrance of the active volume. Thus,
the sampling in timebins along the radial drift allows for a more
refined separation of electrons and pions by exploiting this
information. The simplest extension is a two-dimensional likelihood,
which uses the charges in two time windows as input. As
generalization, the signal is subdivided into 7~slices, which can be
used for higher dimensional methods or as input for a neural network
which is trained for the identification of electrons. The performance
of the methods can be judged by the fraction of pions which pass the
electron cuts for a given electron efficiency. Figure~\ref{fig:eid}
shows a comparison of the different methods. The performance improves
with each TRD layer contributing to the measurement and deteriorates
with increasing momenta since the separation in specific energy loss
decreases and the production of transition radiation saturates.

\begin{figure}
  \centering
  \captionsetup{type=figure}
  \subfloat[jet trigger]{
      \includegraphics[width=.46\textwidth]{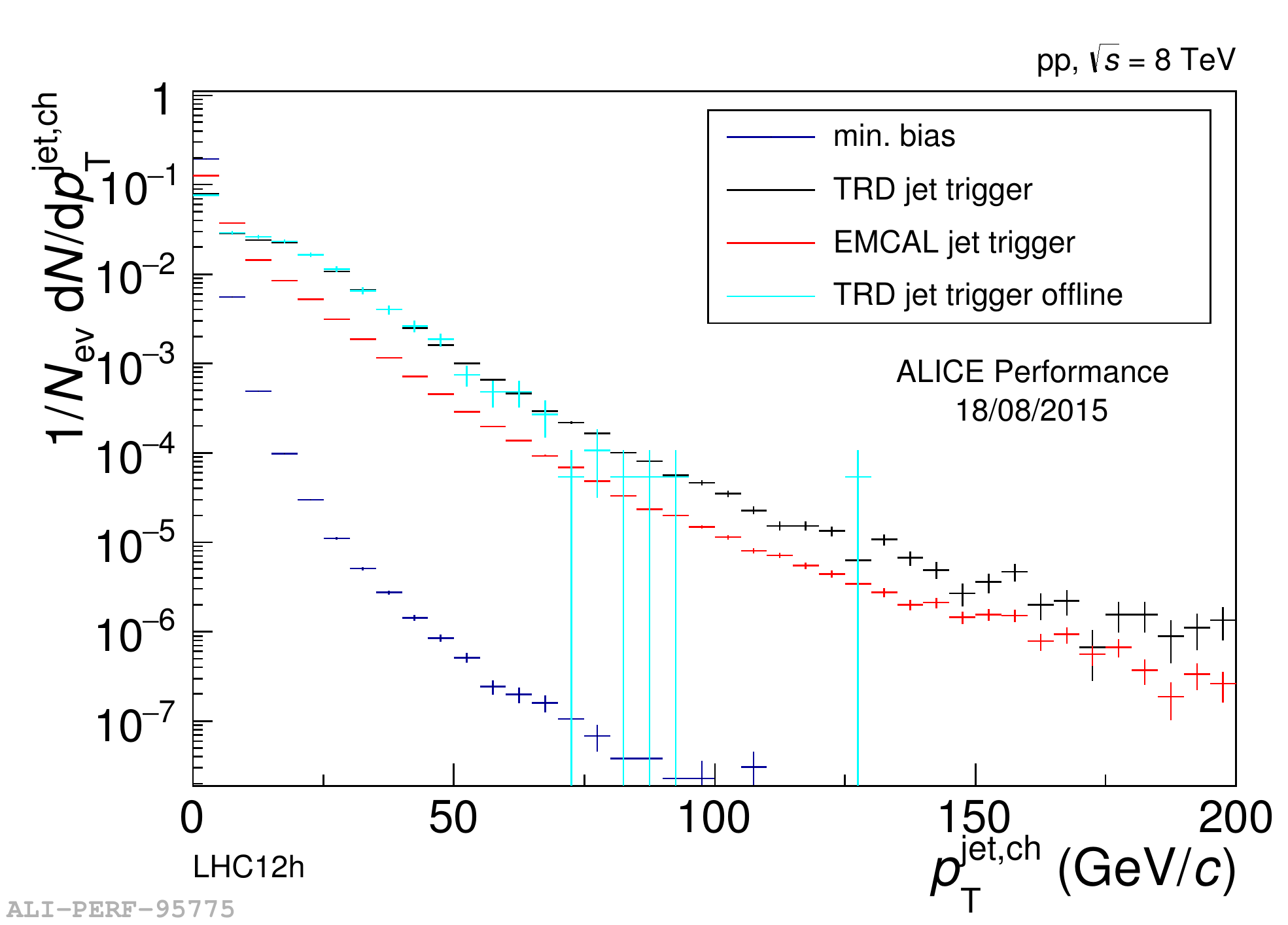}
  }
  \hspace{.5cm}
  \subfloat[electron trigger]{
      \includegraphics[width=.46\textwidth]{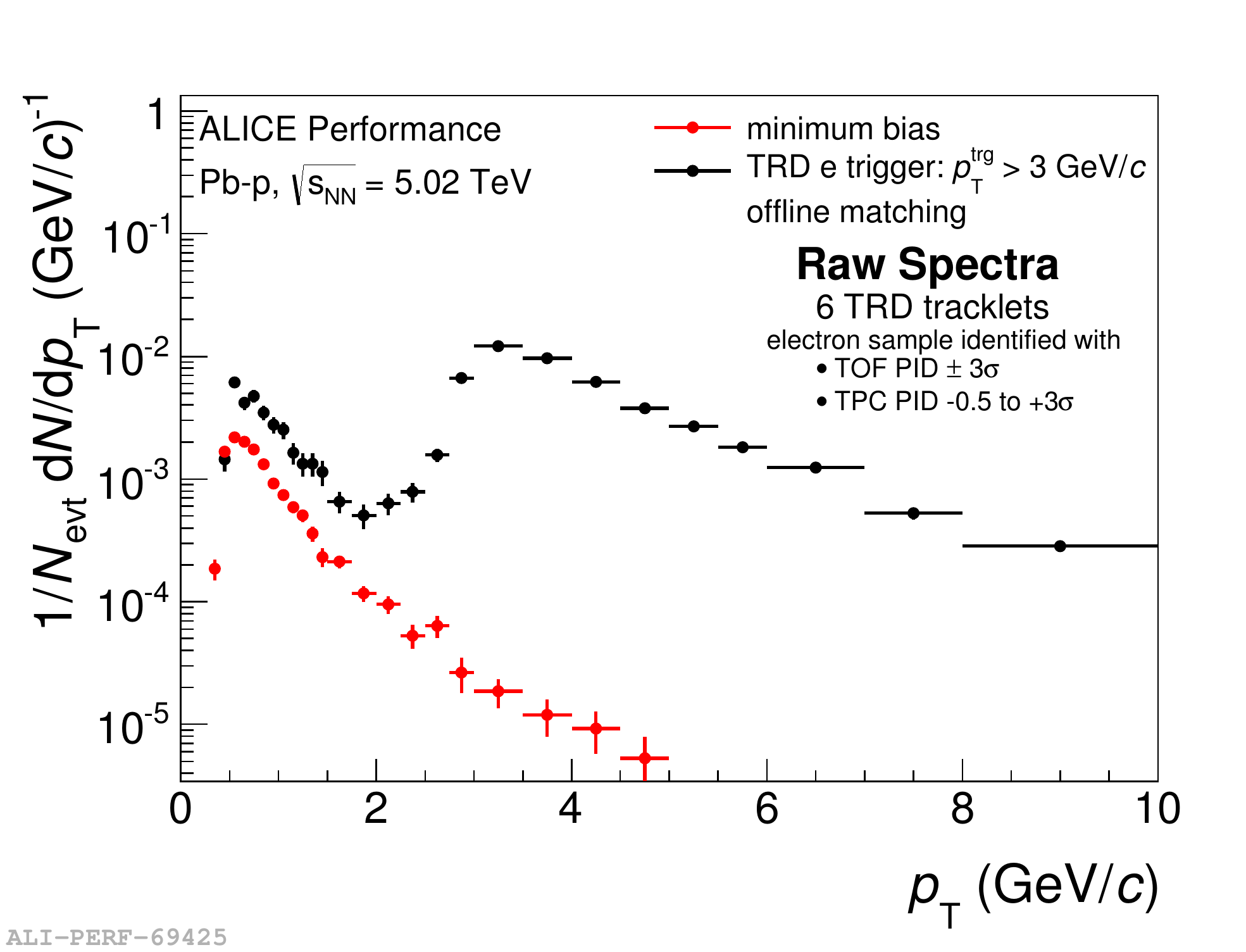}
  }
  \caption{Transverse momentum spectra in TRD-triggered and minimum
    bias samples. Left: We compare the $\pt$ spectra for charged jets
    in the triggered and minimum bias data samples. Right: We compare
    the $\pt$ spectra for electrons in the triggered and minimum bias
    data samples.}
  \label{fig:trigger}
\end{figure}

Besides the usage for the offline reconstruction, the data are used
for the derivation of several contributions to the level-1 trigger of
the experiment, which is issued $6.5~\mu\mathrm{s}$ after the level-0
trigger~\cite{proc:trd:bari}. The low latency poses significant
challenges on the calculations and requires highly parallelized
front-end electronics for local processing. On 256~chamber-mounted
multi-chip modules, each of which comprises an analog pre-amplifier
and shaper and a digital chip with hardware units and four CPUs, the
information is combined into chamber-wise track segments (tracklets).
They are shipped to the Global Tracking Unit (GTU), which combines
them to tracks using an algorithm specifically designed for linear
scaling with multiplicity. For the found tracks, the transverse
momentum $\pt$ and position are calculated. The online tracks serve as
input for a versatile trigger logic, which allows the implementation
of various signatures. Noting that the $\eta$-$\varphi$ coverage of a
TRD stack is comparable to the area of a typical jet cone ($R = 0.2$),
a jet trigger can be derived using the condition that at least three
tracks with $\pt > 3~\mathrm{GeV}/c$ are found in any TRD stack. Even
though only charged particles enter the calculation, the trigger
becomes fully efficient for charged jets with $\pt \gtrsim
100~\mathrm{GeV}/c$. The information on the deposited charge is also
available online and was used for two electron triggers with $\pt$
thresholds of $2$ and $3~\mathrm{GeV}/c$, respectively. The
identification was based on look-up tables translating the total
charge to an electron likelihood. Figure~\ref{fig:trigger} shows the
enhancement by the TRD trigger for jets and electrons.

In addition to extending the physics range by triggering, the TRD has
been used for analyses requiring good and clean identification of
electrons. A prime example is the dielectron decay channel of the
$J/\psi \to e^+ e^-$, for which the TRD helps to improve the
significance of the measurement~\cite{pap:alice_perf}. Also for the
measurement of heavy-flavour mesons through their semi-leptonic decay
channel, the TRD electron identification is used~\cite{Abelev:2012xe}.

\section{LONG SHUTDOWN I}

During the long shutdown of the LHC from 2013 to 2014, the production
of the TRD electronics was completed. This allowed us to finish the
assembly of the five remaining supermodules. They were installed at
the end of 2014 before the experiment was prepared for the start of
\runii{}. The completion of the TRD forms an important milestone in
the project since it allows homogeneous usage of the TRD information
in the full acceptance of the central barrel.

Besides the completion of the detector, consolidation and upgrade
activities were carried out. Some low voltage connections at the
supermodules had shown high resistance which resulted in increased
temperatures and required short-term repairs. The affected
supermodules were removed from the experiment one by one such that the
connections could be reworked in the cavern before the supermodule was
reinstalled. The rework resulted in stable operation of the low
voltage connections for all supermodules.

Since Ethernet is used for the slow control of most detector
components, failures of network components outside of the detector had
resulted in the loss of control over parts of the detector during
\runi{}. Therefore, special multiplexers were developed and
manufactured to realize a redundant connection of the detector
components to the upstream network. The installation for the most
critical components in the supermodules had begun already in \runi{}.
It was completed during the long shutdown such that now all
connections can be remotely switched between two separate uplinks.

The front-end electronics of the TRD requires a wake-up signal prior
to the experiment-wide level-0 trigger. In \runi{}, this signal was
provided by a dedicated pretrigger system. To achieve the required low
latency, it was installed inside the solenoid magnet and received
direct copies of the signals from the trigger detectors. The system
had some limitations in the interoperability with the central trigger
processor, e.g. synchronized down-scaling was not possible. For
\runii{}, the system has been merged with the central trigger
processor to allow for a consistent trigger logic in one place.

\begin{figure}
  \centering
  \captionsetup{type=figure}
  \subfloat[wake-up generation]{
      \includegraphics[height=.45\textwidth]{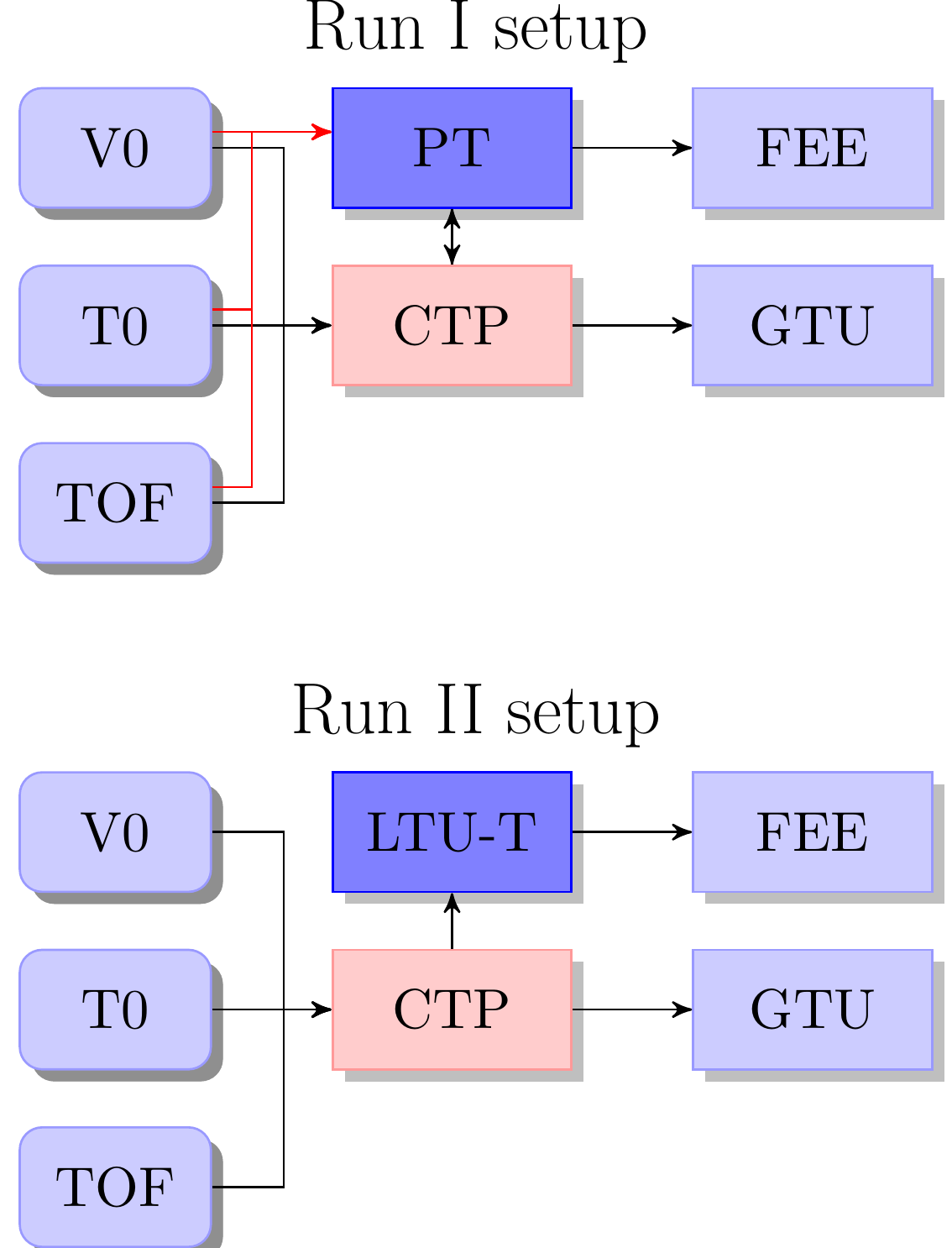}
  }
  \hspace{1cm}
  \subfloat[read-out]{
      \includegraphics[height=.45\textwidth]{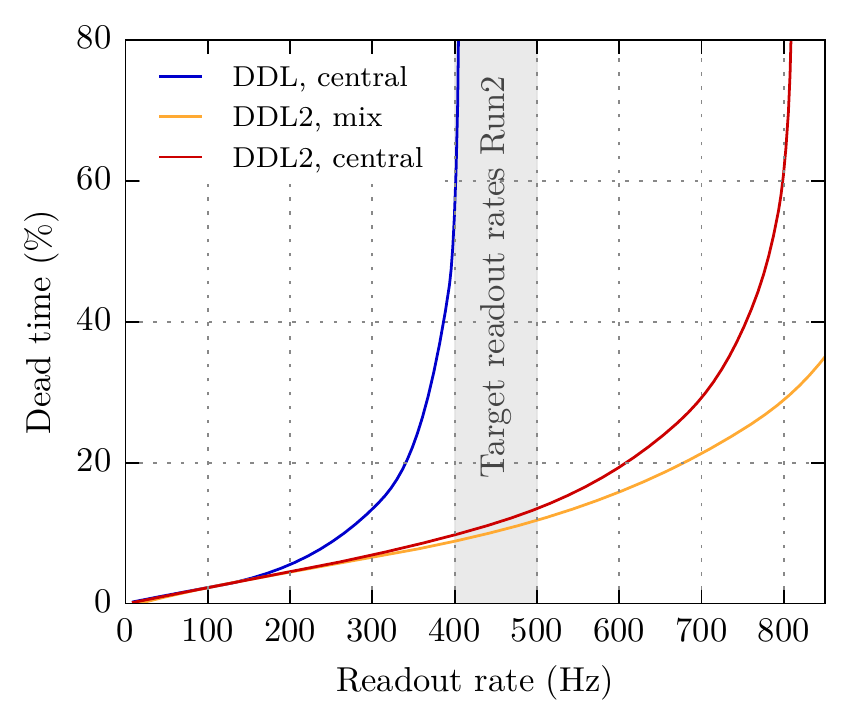}
  }
  \caption{Trigger and read-out upgrade. Left: We show the setup to
    derive the wake-up signal for the TRD front-end electronics in
    \runi{} and \runii{}. Right: We show the dead time corresponding
    to a given read-out rate~\cite{th:skirsch}.}
  \label{fig:readout}
\end{figure}

To avoid a bottleneck in the readout, the Detector Data Links (DDL) to
the data acquisition system were upgraded. The interface logic for the
DDL was migrated from a dedicated mezzanine board to the fabric of the
FPGAs in the GTU, using the in-FPGA multi-gigabit transceivers. This
allowed for a doubling of the read-out bandwidth, resultin in a dead
time similar to \runi{} despite the increased readout rates in
\runii{}, see Figure~\ref{fig:readout}.

\section{RUN II}

After the completion of the detector, the recommissioning started in
the beginning of 2015. The upgraded read-out systems performed as
expected. At first, the full detector was calibrated using Krypton
injected to the gas system~\cite{Stiller:2013zba}. The detector was
fully aligned using early data. After further tweaking of the
detectors used for the wake-up trigger, the upgraded system was
confirmed to fulfill the latency requirements.

\begin{figure}[h]
  \centering
  \captionsetup{type=figure}
  \subfloat[$\pt$ resolution]{
    \begin{minipage}[c][1\width]{.43\textwidth}
      \centering
      \includegraphics[height=\textwidth]{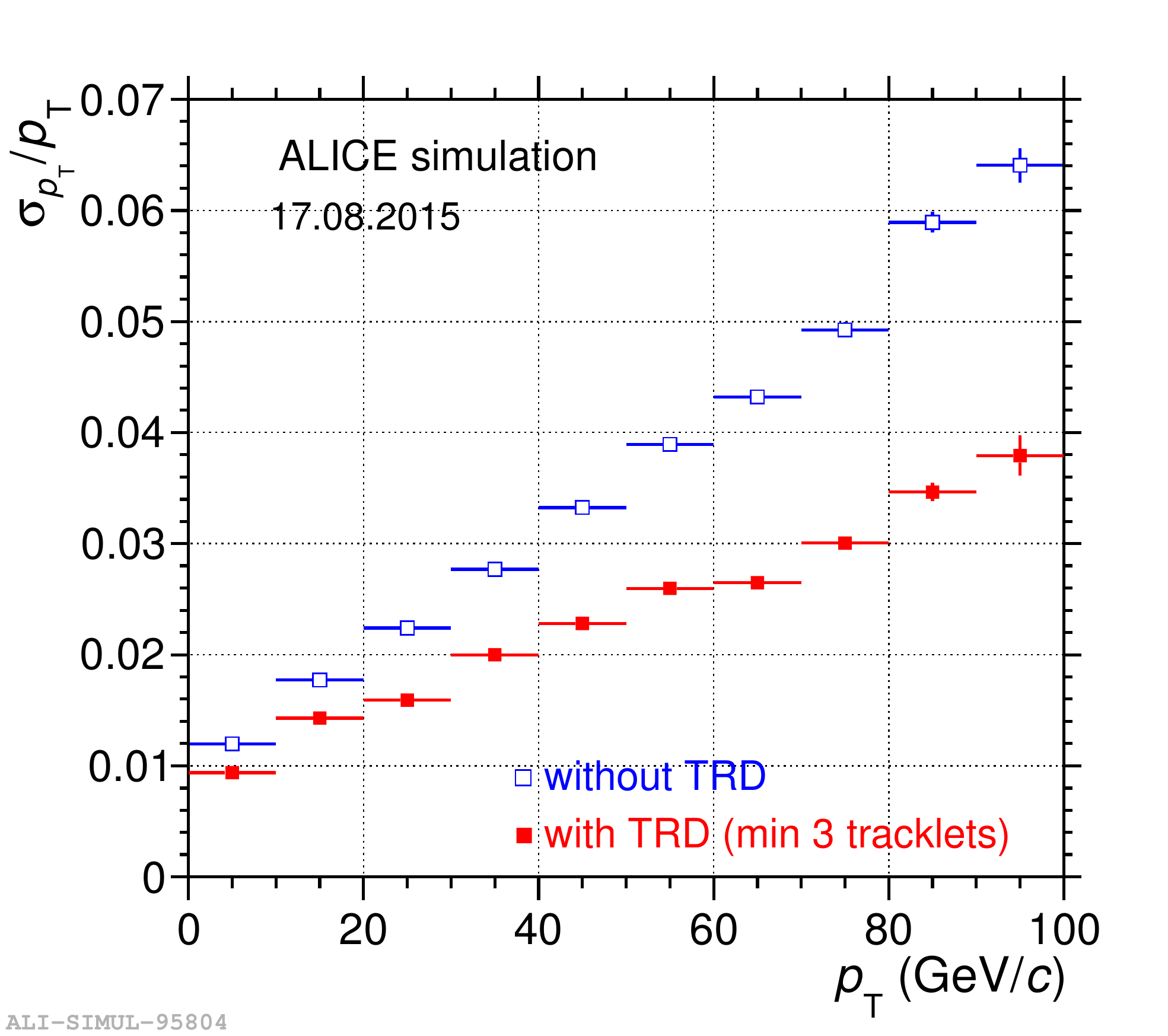}
    \end{minipage}
  }
  \hspace{1cm}
  \subfloat[late conversions]{
    \begin{minipage}[c][1\width]{.43\textwidth}
      \centering
      \includegraphics[height=\textwidth]{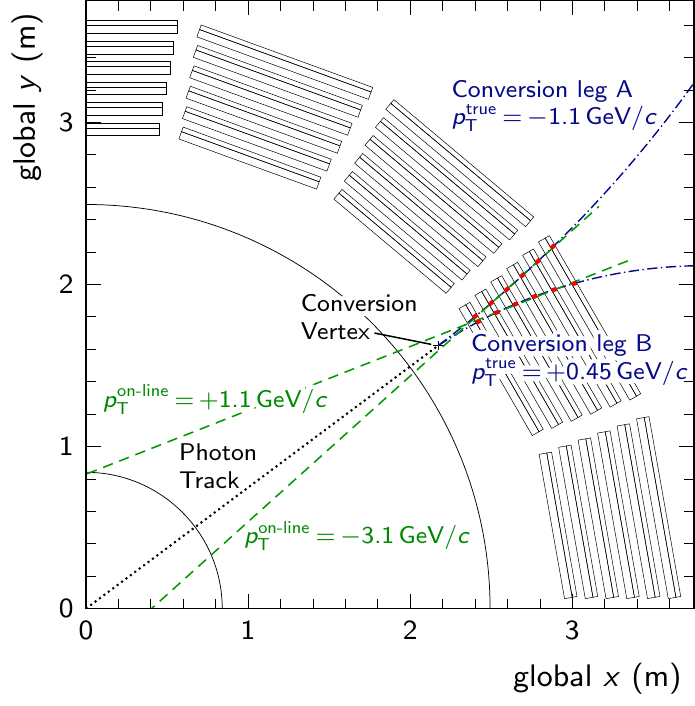}
    \end{minipage}
  }
  \caption{TRD tracking. Left: Transverse momentum resolution without
    and with the TRD used for updating the global track parameters.
    Right: Event display of an exemplary late conversion which wrongly
    fires the trigger.}
  \label{fig:ptres_conv}
\end{figure}

With the full azimuthal coverage of the central barrel acceptance, the
TRD information shall be used to update the track parametrization
during the Kalman propagation. This leads to a significantly improved
$\pt$ resolution, see Figure~\ref{fig:ptres_conv}.

The dominant background for the single electron triggers was caused by
photon conversions at large radii, close to or in the TRD, see
Figure~\ref{fig:ptres_conv}. For the online tracking, the resulting
tracks resemble those with large transverse momenta. For \runii{}, a
rejection of these late conversions has been included in the
FPGA-based online tracking. It compares the $\pt^{-1}$ estimated from
the sagitta and from the global fit and rejects those with a large
discrepancy.


In addition to the electron identification, the TRD can also be used
for hadron identification. For this purpose, the truncated mean is
calculated based on the TRD clusters attached to the track. For the
identification, the deviation from the expectation for a given species
is used after normalization to the resolution expected for the track
under study.

\section{SUMMARY AND OUTLOOK}

The TRD has been completed in time for the start of \runii{} and
performs well now. It shall contribute to the physics output of the
experiment in various areas. The trigger helps to extend the $\pt$
reach for jets and heavy-flavour electrons. The particle
identification is used for analyses of heavy-flavour electrons with
respect to their nuclear suppression factor and the second harmonic
$v_2$.

\bibliographystyle{jkl}
\bibliography{ref}

\end{document}